\documentstyle[prl,aps]{revtex}
\twocolumn
\begin{document}
\author{Jian Qi Shen\footnote{E-mail address: jqshen@coer.zju.edu.cn}}
\address{Centre for Optical
and Electromagnetic Research, Zhejiang University, Hangzhou Yuquan
310027, P.R. China}
\date{\today }
\title{Comment on ``Quantum vacuum contribution to the momentum of dielectric media''}
\maketitle

\begin{abstract}
PACS number(s): 42.50.Lc, 78.20.Ci, 42.50.Nn
\end{abstract}

Some novel vacuum effects associated with the dynamical quantities
(such as energy, spin, polarization and momentum) of quantum
vacuum fluctuation fields have been considered in the
literature\cite{Casimir,Lam,Rikken,Fuentes,Shenpla,Shen,Feigel}.
These effects include the Casimir effect (due to the change of the
vacuum mode structure)\cite{Casimir,Lam}, magnetoelectric
birefringences of the quantum vacuum\cite{Rikken}, vacuum-induced
Berry's phase of spinning particles (caused by the interaction
between magnetic moments and vacuum zero-point
energies)\cite{Fuentes} and quantum-vacuum geometric phase of
zero-point field in a coiled fiber system (related to the spin and
polarization of the vacuum fluctuation energy)\cite{Shenpla,Shen}.
As to the effect associated with the momentum of vacuum zero-point
fields, more recently, Feigel has considered the quantum vacuum
contribution to the momentum of electromagnetic
media\cite{Feigel}. However, in Feigel's treatment he did not take
into account the relativistic transformation of the optical
constants (electric permittivity and magnetic permeability) of
moving media. We think that it is necessary to attach importance
to the relativistic transformation of the optical constants in
this subject. Here we will show that the effect arising from such
a transformation will also provide a quantum vacuum contribution
to the velocity of media, in addition to the one derived by Feigel
himself\cite{Feigel}.

Consider an electromagnetic medium moving along the $\hat{{\bf
 z}}$-direction at a velocity ${\bf v}$ relative
to a rest frame of reference, K, the relativistic transformations
of $\epsilon$ and $\mu$ (the components in the $\hat{{\bf
 x}}-\hat{{\bf
 y}}$ plane) take the form %\cite{Shensr}
\begin{equation}
\mu'=\sqrt{\frac{\mu}{\epsilon}}\left(\frac{\sqrt{\epsilon\mu}+\frac{v}{c}}{1+\sqrt{\epsilon\mu}\frac{v}{c}}\right),
   \quad
\epsilon'=\sqrt{\frac{\epsilon}{\mu}}\left(\frac{\sqrt{\epsilon\mu}
+\frac{v}{c}}{1+\sqrt{\epsilon\mu}\frac{v}{c}}\right),
\end{equation}
where $\mu$ and $\epsilon$ are the intrinsic permeability and
permittivity of the medium, respectively, and $\mu'$ and
$\epsilon'$ the permeability and permittivity (observed from the
rest frame K) of this moving medium.

As stated by Feigel, in the case of magnetoelectrics, a term
$\left({1}/{\mu}\right){\bf B}\cdot\hat{\chi}^{\rm T}{\bf E}$ must
be added to the Lagrangian of the electromagnetic
system\cite{Feigel}. Such a term in a moving medium can be
expanded up to the first order in $v/c$, {\it i.e.},
\begin{eqnarray}
&  &  \frac{1}{\mu}{\bf B}\cdot\hat{\chi}^{\rm T}{\bf
E}+\frac{1}{\mu c}\left[{\bf B}\cdot\hat{\chi}^{\rm T}\left({\bf
v}\times {\bf B}\right)+\left({\bf E}\times
{\bf v}\right)\cdot\hat{\chi}^{\rm T}{\bf E}\right]     \nonumber \\
& & +\frac{1}{\mu
c}v\left(\sqrt{\epsilon\mu}-\frac{1}{\sqrt{\epsilon\mu}}\right){\bf
B}\cdot\hat{\chi}^{\rm T}{\bf E}+{\mathcal
O}\left(\frac{v^{2}}{c^{2}}\right). \label{eqeq37}
\end{eqnarray}
Compared with the result derived by Feigel, the expression
$\left({1}/{\mu
c}\right)v\left[\sqrt{\epsilon\mu}-\left({1}/{\sqrt{\epsilon\mu}}\right)\right]{\bf
B}\cdot\hat{\chi}^{\rm T}{\bf E}$ in (\ref{eqeq37}) is a new term,
which arises from the relativistic transformation of $\mu$. Since
the velocity ${\bf v}$ of the medium is parallel to the $\hat{{\bf
 z}}$-direction, {\it i.e.}, ${\bf v}=v\hat{{\bf
 z}}$ with $\hat{{\bf
 z}}$ being a unit vector, the Lagrangian of the moving magnetoelectric medium is
 thus of the form
\begin{eqnarray}
L_{\rm ME}&=& L_{FM}+\int\frac{{\rm
d}^{3}x}{4\pi}\left(\frac{1}{\mu}{\bf B}\cdot\hat{\chi}^{\rm
T}{\bf E}\right)
    \nonumber   \\
&&  +\frac{1}{\mu c}\int\frac{{\rm d}^{3}x}{4\pi}{\bf
v}\cdot\left\{\left[{\bf B}\times\left(\hat{\chi}{\bf
B}\right)\right]-\left[{\bf E}\times\left(\hat{\chi}^{\rm T}{\bf
E}\right)\right]\right\}
    \nonumber   \\
& & + \frac{1}{\mu c}\int\frac{{\rm d}^{3}x}{4\pi}{\bf
v}\cdot\hat{{\bf
z}}\left(\sqrt{\epsilon\mu}-\frac{1}{\sqrt{\epsilon\mu}}\right){\bf
B}\cdot\hat{\chi}^{\rm T}{\bf E}.   \label{totlag}
\end{eqnarray}
For the definition of $L_{FM}$, see Eq. (9) in Ref.\cite{Feigel}.
Note that the final term on the right-handed side of Eq.
(\ref{totlag}) in the present Comment is new compared with Eq.
(19) in Feigel's paper\cite{Feigel}. Thus, according to the
Lagrange equation (liquid's equation) of motion\cite{Feigel}, one
can obtain the following equation
\begin{eqnarray}
\rho^{0}v\hat{{\bf z}}&=&\frac{1}{4\pi\mu
c}\left[\left({\epsilon\mu-1}\right){\bf E}\times{\bf B}+{\bf
E}\times\left(\hat{\chi}^{\rm T}{\bf E}\right)-{\bf
B}\times\left(\hat{\chi}{\bf B}\right)\right]     \nonumber  \\
& &  -\frac{1}{4\pi\mu
c}\left(\sqrt{\epsilon\mu}-\frac{1}{\sqrt{\epsilon\mu}}\right)\left({\bf
B}\cdot\hat{\chi}^{\rm T}{\bf E}\right)\hat{{\bf z}}.
\label{final}
\end{eqnarray}
Note that the final term on the right-handed side of Eq.
(\ref{final}) is a new quantum vacuum contribution to the momentum
of the medium, which has not yet been taken into consideration in
Feigel's work\cite{Feigel}.


\begin{references}
\bibitem{Casimir} H.B.G. Casimir, Proc. K. Ned. Akad. Wet. {\bf
51}, 793 (1948).

\bibitem{Lam} S.K. Lamoreaux, Phys. Rev. Lett. {\bf 78}, 5 (1997).

\bibitem{Rikken}  G.L.J.A. Rikken and C. Rizzo, Phys. Rev. A {\bf
63}, 012107 (2000).


\bibitem{Fuentes}    I. Fuentes-Guridi, A. Carollo, S. Bose, and V. Vedral, Phys. Rev. Lett. {\bf 89},
220404 (2002).

\bibitem{Shenpla}  J.Q. Shen and L.H. Ma, Phys. Lett. A {\bf 308},
355 (2003).


\bibitem{Shen} J.Q. Shen, J. Opt. B: Quantum Semiclass. Opt. {\bf
6}, L13 (2004).

\bibitem{Feigel}  A. Feigel, Phys. Rev. Lett. {\bf 92}, 020404 (2004).

%\bibitem{Shensr} J.Q. Shen, arXiv: physics/0408005 (2004).

\end{references}
\end{document}